\documentclass[structabstract]{aa}  
\usepackage{graphicx}
\usepackage{txfonts}
\begin{document}
   \title{Integral field optical spectroscopy of a representative sample of ULIRGs.}

   \subtitle{II. Two-dimensional kpc-scale extinction structure\thanks{Based on observations with the William Herschel Telescope
  operated on the island of La Palma by the ING in the Spanish Observatorio
  del Roque de los Muchachos of the Instituto de Astrof\'{\i}sica de
  Canarias. Based also on observations with the NASA-ESA Hubble Space
  Telescope, obtained at the Space Telescope and Science Institute, which is
  operated by the Association of Universities for Research in Astronomy,
  Inc. under NASA contract number NAS5-26555.}}

   \author{M. Garc\'{\i}a-Mar\'{\i}n
          \inst{1, 2}
          \and
          L. Colina\inst{1}
          \and
          S. Arribas\inst{1}
          }
\offprints{M. Garc\'{\i}a-Mar\'{\i}n (maca@ph1.uni-koeln.de)}
   \institute{Departamento de Astrof\'{\i}sica Molecular e Infrarroja, Instituto de Estructura de la Materia, CSIC
              C/ Serrano 121, Madrid, Spain\\
         \and
             Present address: I. Physikalisches Institut, Universit\"at zu K\"oln
Z\"ulpicher Strasse 77, 50937 K\"oln, Germany\\
             \email{maca@ph1.uni-koeln.de}
}

   \date{Received ; accepted}

 
  \abstract
   {Ultraluminous infrared galaxies are merging systems characterized for containing large amounts of dust and emitting the bulk of their energy in the infrared.
Dust affects several of the observed properties derived from optical and near-IR rest-frame data such as the stellar morphology, and the ionized gas distribution. Systematic analysis of the dust distribution in representative samples of ULIRGs are needed to investigate its two-dimensional structure, and to establish its impact in the derivation of fundamental properties such as star formation rates, effective radii, and dynamical masses. 
}
   {We investigate the two-dimensional kpc-scale structure of the extinction
    in a representative sample of local ULIRGs using the H$\alpha$/H$\beta$ line ratio.  
}
   {We use optical integral field spectroscopy obtained with the INTEGRAL instrument at the William Herschel Telescope. Complementary optical and near-IR high angular resolution HST images have also been used. 
Compared to classical optical long-slit analysis, our data provides information about the entire system, and not only in pre-selected orientations. 
}
   {The extinction exhibits a very complex and patchy structure in ULIRGs on kpc scales, from basically transparent regions to others deeply embedded in dust ({$A_{\rm{V}}$})$\simeq$0.0 to {$A_{\rm{V}}$}$\simeq$8.0 mag). Nuclear extinction covers a broad range in {$A_{\rm{V}}$} from 0.6 to 6 mag, 69\% of the nuclei having $A_{\rm{V}}$$\ge$2.0 mag. Extinction in the external regions is substantially lower than in the nuclei with 64\% of the ULIRGs in the sample having median {$A_{\rm{V}}$} of less than 2 mag for the entire galaxy.
While post-coalescence nuclei tend to cluster around {$A_{\rm{V}}$} values of 2 to 3 mag, pre-coalescence nuclei appear more homogeneously distributed over the entire 0.4 mag {$\leq$$A_{\rm{V}}$ $\leq$} 7.7 mag range. 
For the average extinction ({$A_{\rm{V}}$$\simeq$2.0}) derived for the ULIRGs of the sample, the ratio of the de-reddened to observed \textit{SFR} values is 6.  The extinction-corrected, H$\alpha$-based \textit{SFR} ranges from 10 to 300 M$\odot$ yr$^{-1}$. For only 28\% of the cases the de-reddened \textit{SFR} is $<$20 M$\odot$ yr$^{-1}$, whereas for the observed \textit{SFR} this percentage increases to 72\%. The IR-based \textit{SFR} is always higher than the optical-based one, with differences ranging from about 2 to up to 30. The nuclear observed \textit{SFR} has an average contribution to the total one of 16\% for the entire sample. Once corrected for extinction, the average value becomes 31\%. 

Because of mostly extinction effects, the optical (I-band) half-light radius 
in the sample galaxies is on average a factor 2.3 larger than the corresponding near-IR (H-band) value. 
}
   {}


   \titlerunning{IFS of ULIRGs. II. The Dust Structure}
   \authorrunning{Garc\'{\i}a-Mar\'{\i}n et al.}  

   \maketitle
%

\section{Introduction}

The significance of galaxies with an infrared excess was highlighted early by Low \& Kleinmann (1968) and Rieke \& Low (1972), and lately recognized by observations of the IRAS satellite (Soifer et al. 1984). Among the sources that emit the bulk of their energy in the infrared, the Ultra-Luminous Infrared Galaxies (ULIRGs, {10$^{12}$$L_{\odot}$$\le$ $L_{\rm{bol}}$$\sim$ $L_{\rm{IR}}$$[8-1000\mu\rm{m}]$ $\le$ 10$^{13}$$L_{\odot}$}), are amongst the brightest objects in the local Universe (see reviews in Lonsdale et al.\,2006; Sanders \& Mirabel 1996). The process of merging with accompanying starbursts plays a key role in these galaxies, and appears to transform spirals into low-intermediate mass ellipticals (e.g., Colina et al.\,2001; Genzel et al.\,2001; Tacconi et al.\,2002; Naab et al.\,2006 and references therein). This star formation activity is the main energy source in ULIRGs, although in some cases the contribution of an obscured AGN is relevant (see Nardini et al.\,2008; Risaliti et al.\,2006). These galaxies contain large quantities of gas and dust that,
independent of the nature of the energy source, are heated re-emitting most of their energy in the infrared. 

High angular resolution {\textit{HST}} optical and near-infrared images (Farrah et al. 2001; Bushouse et al. 2002) infer that ULIRGs have in general a complex stellar structure. On the one hand, these structures are caused by the merger itself that transforms the stellar distribution in the parent galaxies (i.e., into tidal tails, nuclear and extended star-forming regions, and double nuclei) as they evolve through the different phases of the interaction. On the other hand, large amounts of gas and dust settle into extended lanes and/or filaments creating additional structures due to non-uniform extinction effects (see V, I, and H band images in Farrah et al. 2001 and Bushouse et al. 2002). The impact of the dust effects is lower in the near-IR, allowing us to identify the true stellar morphology of the galaxies (e.g., {\textit{HST}} observations from Scoville et al. \, 2000 and Bushouse et al.\,2002). This is emphasized by studies based on {\textit{HST}} color maps (F814W-F606W), which detect a non uniform color structure with blue and red compact knots that are most likely caused by unobscured star formation, and dust-enshrouded star formation or remnant nuclei from the progenitor galaxies (Surace et al. \, 1998; Farrah et al.\,2001). Similar results have been obtained for luminous infrared galaxies (LIRGs, {10$^{11}$$L_{\odot}$$\le$ $L_{\rm{bol}}$$\sim$ $L_{\rm{IR}}$$[8-1000\mu\rm{m}]$ $\le$ 10$^{12}$$L_{\odot}$}) using near-IR {\textit{HST}} images. Complex dust features in their centers, with an average optical extinction ({{$A_{\rm{V}}$}}) of between 3 and 5 mag, are also observed (Alonso-Herrero et al.\,2006).

 All the afore mentioned \textit{HST} imaging studies indicate that although the dust tends to be concentrated in the inner few kpc, the global distribution is very patchy on scales of hundreds to thousands of parsecs, and therefore non-uniform extinction effects play a major role. ULIRGs have been intensively studied with narrow, long-slit spectroscopy (e.g., Veilleux et al.\,1995), but despite its importance, the spectroscopic derivation of the two-dimensional dust and gas distribution has not been the subject of much scrutiny. Previous integral field spectroscopy (IFS) studies of specific targets demonstrated the lack of coincidence between the true, dynamical nucleus of the galaxy and the region identified as the optical nucleus, which can be separated by distances from 0.5 (Arp~299, Garc\'{\i}a-Mar\'{\i}n et al.\,2006) to 1.3 kpc (IRAS~17208$-$0014, Arribas \& Colina 2003) . Other studies have also shown the large differences (of up to a factor of 10) in extinction between the nuclear, circum-nuclear, and extranuclear regions due to the patchy and non-uniform distribution of the dust on kpc-scales (e.g. IRAS 12112$-$0305, Colina et al. \, 2000; Arp~299, Garc\'{\i}a-Mar\'{\i}n et al.\,2006). These results emphasize the importance of knowing (and correcting for) the two-dimensional structure of the extinction before using rest-frame optical lines (mainly H$\alpha$) or structural parameters (e.g., effective radius) in the derivation of star formation rates and dynamical masses in these galaxies, and their intermediate- and high-z analogs.

This is the second paper. in a series aimed at studying in detail the internal structure and kinematics of local ULIRGs (Garc\'{\i}a-Mar\'{\i}n et al.\,2009, Paper I). Here, we present the first systematic two-dimensional analysis of the dust extinction in a representative sample of low-{\textit{z}} ULIRGs using optical integral field spectroscopy complemented with existing archival \textit{HST} high angular resolution images. The paper is organized as follows. The galaxy sample is presented in Sect. 2, whereas Sect. 3 is dedicated to the observations and data reduction, and Sect. 4 presents the data analysis. Section 5 discusses the extinction structure of ULIRGs on kpc scales, and the implications for the derivation of star formation rates and dynamical masses in low- and high$-z$ ULIRGs. Finally, in Sect. 6, a brief summary of the main results is given.

The current analysis that we are carrying out in this sample of ULIRGs is part of a larger survey that is investigating the two-dimensional extinction, ionization, and kinematic kpc-scale structure of a representative sample of low$-z$ LIRGs and ULIRGs, and its implications for their high-redshift analogs. The survey is based on the use of optical IFS data obtained with different facilities.  The study of northern hemisphere ULIRGs has been performed mostly with the integral field unit (IFU) INTEGRAL (Arribas et al.\,1998). Several papers presenting the results for individual (Colina, Arribas \& Borne, 1999; Garc\'{\i}a-Mar\'{\i}n et al.\,2006) or small subsamples of galaxies (see Colina et al.\,2005; Monreal-Ibero et al.\,2007 and references therein) have already been published.

To include southern systems and extend the sample to lower luminosity objects covering most of the  (U)LIRGs range (log($L_{\rm{IR}}$/$L\odot$) = 11.0-12.6), we observed an additional sample of about 40 systems using VIMOS IFU (LeFevre et al.\,2003). The VIMOS sample description and first results were presented in Arribas et al.\,(2008).

We also observed the majority of the northern hemisphere galaxies from the volume-limited (distances of about 35-75~Mpc) local sample of LIRGs presented in Alonso-Herrero et al.\,(2006); further details of the selection criteria can be found in Alonso-Herrero et al.\, 2006. These galaxies, with an average luminosity of log($L_{\rm{IR}}$/$L\odot$)=11.32, were observed with the IFU PMAS (Roth et al.\,2005). The data and first results are presented in Alonso-Herrero et al.\,(2009, Paper I.).

In addition to this, near-IR IFS with SINFONI of a subsample of LIRGs and ULIRGs is also being conducted (see Bedregal et al.\,2009 for first results) to investigate the multiwavelength (optical and near-IR) properties of these galaxies.

Throughout this paper, we use $\Omega_{\Lambda}$=0.7, $\Omega_{M}$=0.3, and $H_{0}$=70 km s$^{-1}$ Mpc$^{-1}$.


\section{The sample of ULIRGs}

Our original sample of ULIRGs with available optical INTEGRAL IFS consists of 22 systems selected to be representative of this galaxy class (details on the sample selection can be found in Garc\'{\i}a-Mar\'{\i}n et al.\, 2009, Paper I). These systems exhibit a large variety of morphologies, from wide pairs to close pairs with well developed tidal tails to single nucleus mergers. The sample covers the luminosity range 11.8$\le$log{{($L_{\rm{IR}}$/$L_{\odot}$)}} $\le$12.6, and includes all the different classes of nuclear activity (H\,{\sc{ii}}-, LINER-, and Seyfert-like). A few galaxies of the sample are not present in the IFS-based analysis of the dust distribution for a variety of reasons, such as the lack of detection or non coverage of H$\beta$ (Arp~220 and IRAS~09427+1929), low S/N, which precludes the detection of emission lines (IRAS~13469+5833), limited spatial resolution (IRAS~13342+3932), and severe AGN contamination (Mrk~231). Using the projected distance between the nuclei of the parent galaxies, the sample has been further divided into two categories, pre- (nuclear distance $>$1.5~kpc) and post-coalescence (nuclear distance $\leq$1.5~kpc). This separation allows us to investigate and discriminate the characteristics of galaxies in the final phases of the merger from those in earlier dynamical stages. Given this, the two-dimensional dust distribution has been derived for 17 (U)LIRGs, nine of them pre-coalescence systems and eight post-coalescence (Table \ref{Table1}).

\begin{table*}[t]
\begin{minipage}[t]{\textwidth}
\centering
\renewcommand{\footnoterule}{}  
\caption{Extinction and structural properties for the sample of galaxies.}   
\label{Table1}      
\tabcolsep0.1cm
\begin{tabular}{lccccccccccc}       
\noalign{\smallskip}
\hline
\noalign{\smallskip}
\hline
\noalign{\smallskip}
Galaxy\footnote{The galaxies are listed according to decreasing nuclear separation (see Garc\'{\i}a-Mar\'{\i}n et al 2009).} &log({$L_{\rm{IR}}$})\footnote{{$L_{\rm{IR}}$}(8-1000 $\mu$m) was derived following Sanders \& Mirabel 1996)} & \textit{z} & {$A_{\rm{V}}$$_{\rm{median}}$}\footnote{Median extinction of the entire system with the deviation that effectively represents the distribution of the extinction values over the entire area where the H$\alpha$ and H$\beta$ measurements can be obtained. For pre-coalescence systems with one pointing per nuclei, the median has been calculated for each pointing.} & {$A_{\rm{V}}$$_{\rm{nuc}}$}\footnote{Nuclear extinction of the galaxy. The uncertainties are about 20\%} & {{$A_{\rm{V}}$$_{\rm{max}}$}}\footnote{Maximum extinction of the system.} &Sep\footnote{Separation between the optical nucleus, identified as the peak position in the F814W band of the HST images, and the position of the peak of extinction.} & Scale & {{$R_{\rm{eff}}$}}(optical) & {{$R_{\rm{eff}}$}}\footnote{Effective radius measured from the HST WFPC2 camera, with the filter F814W, corresponding to a central wavelength of 0.82\,$\mu$m and a filter width of 0.17\,$\mu$m.} & {{$R_{\rm{eff}}$}}\footnote{Effective radius measured from the HST NICMOS2 camera, with the filter F160W, corresponding to a central wavelength of 1.55\,$\mu$m and a filter width of 0.40\,$\mu$m.} & Morphology\footnote{IP means interacting pair, DN double nucleus and SN single nucleus.} \\ 
       & {{$L_{\odot}$}} & & (mag) & (mag)   &(mag)     & (kpc) & (kpc/arcsec) & (kpc) & (kpc) & (kpc) & \\
\hline
IRAS~13156+0435N   &12.13 &0.113 &  2.5{$\pm$}0.8  & 3.3 & 3.7 & 3.9 $\pm$1.0 & 2.058  &                     & 5.9 &                        &  IP at 36.0~kpc\\
IRAS~13156+0435S   &      &      &  0.9$\pm$0.4  & 1.5 & 1.7 & 2.2 $\pm$1.0 & 2.058  &                     & 4.9 &                        &                \\
IRAS~18580+6527E   &12.26 &0.176 &  1.3$\pm$0.8  & 0.5 & 3.9 & 20.3$\pm$1.5 & 2.986  & 11.3\footnote{Effective radius measured from the HST WFPC2 camera, with the filter F606W, corresponding to a central wavelength of 0.58\,$\mu$m, and a filter width of 0.16\,$\mu$m.}& 6.5 &    &  IP at 15.0~kpc\\
IRAS~18580+6527W   &      &      &               & 0.8 & 3.9 & 8.7 $\pm$1.5 & 2.986  & 9.4$^{j}$           & 4.3 & &                \\
IRAS~16007+3743E   &12.11 &0.185 &  1.9$\pm$1.0  & 4.5 & 5.0 & 12.7$\pm$1.5 & 3.100  &                     & 4.0 & &  IP at 14.2~kpc\\
IRAS~16007+3743W   &      &      &               & 4.7 & 5.0 & 1.4 $\pm$1.5 & 3.100  &                     & 4.2 & &                \\
IRAS~06268+3509N   &12.51 &0.169 &  2.0$\pm$1.2  & 5.7 & 6.9 & 3.2 $\pm$1.4 & 2.895  & 8.3$^{j}$           & 2.0 & 2.1       & IP at 9.1~kpc \\
IRAS~06268+3509S   &      &      &               & 2.2 & 6.9 & 10.5$\pm$1.4 & 2.895  & 6.3$^{j}$           & 1.5 & 1.1       &                \\
IRAS~08572+3915N   &12.17 &0.058 &  1.7$\pm$1.1  & 2.3 & 5.8 & 0.0 $\pm$0.6 & 1.130  & 2.9\footnote{Effective radius measured from the HST WFPC2 camera, with the filter F439W, corresponding to a central wavelength of 0.43\,$\mu$m and a filter width of 0.05\,$\mu$m.}                 & 2.8 & 0.6  &IP at 6.1~kpc \\
IRAS~08572+3915S   &      &      &               & 0.4 & 5.8 & 1.3 $\pm$0.6 & 1.130  & 4.3$^{k}$           & 1.4 & 0.6  &              \\
IRAS~14348$-$1447N &12.39 &0.083 &  2.9$\pm$1.0  & 5.4 & 5.8 & 5.6 $\pm$0.8 & 1.556  & 5.3\footnote{Effective radius measured from the HST ACS camera, with the filter F435W, corresponding to a central wavelength of 0.43\,$\mu$m and a filter width of 0.14\,$\mu$m.}                & 1.4 & 1.5  &IP at 5.5~kpc \\
IRAS~14348$-$1447S &      &      &               & 4.1 & 5.8 & 0.8 $\pm$0.8 & 1.556  & 4.1$^{l}$           & 3.0 & 1.3  &                \\
Mrk~463-E          &11.81 &0.050 &  0.9$\pm$0.5  & 0.7 & 3.0 & 4.7 $\pm$0.5 & 0.984  &                     & 1.0 & 0.2  &  DN at 3.8~kpc \\
Mrk~463-W          &      &      &               & 1.5 & 3.0 & 8.4 $\pm$0.5 & 0.984  &                     & 1.0 & 1.0  &                \\
Arp~299/NGC~3690   &11.81 &0.010 &  1.9$\pm$0.6  & 3.7 & 4.3 & 0.5 $\pm$0.1 & 0.205  &                     & 1.2 & 0.9  &  IP at 5.0~kpc \\
Arp~299/IC~694     &      &      &  2.0$\pm$0.6  & 2.7 & 3.4 & 0.3 $\pm$0.1 & 0.205  &                     & 2.0 & 0.4  &                \\
IRAS~12112+0305N   &12.37 &0.073 &  1.7$\pm$1.3  & 3.0 & 7.7 & 5.5 $\pm$0.7 & 1.395  & 3.3$^{l}$           & 2.0 & 1.7  &  IP at 4.0~kpc \\
IRAS~12112+0305S   &      &      &               & 7.7 & 7.7 & 0.0 $\pm$0.7 & 1.395  & 4.5$^{l}$           & 1.5 & 1.1  &                 \\
IRAS~06487+2208    &12.57 &0.144 &  2.2$\pm$0.6  & 2.4 & 3.5 & 4.3 $\pm$0.6 & 2.522  &                     & 1.5 & &  DN at 1.5~kpc \\
IRAS~11087+5351    &12.13 &0.143 &  1.2$\pm$1.0  & 3.0 & 3.0 & 0.0 $\pm$0.6 & 2.507  &                     & 8.9 & &  DN at 1.5~kpc \\
Mrk~273            &12.18 &0.038 &  1.9$\pm$1.0  & 3.1 & 4.9 & 2.1 $\pm$0.4 & 0.749  & 4.3$^{l}$           & 4.4 & 2.0  &  DN at 0.7~kpc \\
IRAS~12490$-$1009  &12.07 &0.101 &  1.3$\pm$0.7  & 1.1 & 3.3 & 3.6 $\pm$0.9 & 1.854  &                     & 7.6 & &  SN            \\
IRAS~14060+2919    &12.18 &0.117 &  1.5$\pm$0.5  & 2.2 & 2.5 & 2.3 $\pm$1.0 & 2.113  &                     & 4.8 & &  SN            \\
IRAS~15206+3342    &12.27 &0.124 &  0.8$\pm$0.5  & 1.0 & 2.0 & 2.4 $\pm$1.0 & 2.232  & 2.2$^{k}$           & 2.9 &      &  SN            \\
IRAS~15250+3609    &12.09 &0.055 &  2.3$\pm$1.7  & 2.5 & 7.9 & 2.4 $\pm$0.5 & 1.072  & 2.2$^{l}$           & 1.9 & 1.4  &  SN            \\
IRAS~17208$-$0014  &12.43 &0.043 &  5.5$\pm$2.6  & 5.0 & 9.0 & 0.8 $\pm$0.4 & 0.844  & 3.6$^{l}$           & 3.2 & 1.5  &  SN            \\
\hline                                   
\end{tabular}
\end{minipage}
\end{table*} 

\section{Observations and data reduction}
IFS data of the galaxy sample was obtained between 1998 and 2004 using INTEGRAL, a fiber-based optical integral field system (Arribas et al.\,1998) connected to the Wide Field Fibre Optic Spectrograph (WYFFOS; Bingham et al.\,1994) and mounted on the 4.2 m William Herschel Telescope. Depending on the structure and compactness of the ULIRGs,three different INTEGRAL configurations were used: the so-called standard bundles 1, 2, and 3 (SB1 with fiber diameter 0.45 arcsec and field-of-view (FoV) 7.8$\times$ 6.4 arcsec$^{2}$, SB2 with fiber diameter 0.9 arcsec and FoV 16.0$\times$ 12.3 arcsec$^{2}$, and SB3 with fiber diameter 2.7 arcsec and FoV 33.6$\times$ 29.4 arcsec$^{2}$). The spectra were taken with a 600 lines mm$^{-1}$ grating, providing an effective spectral resolution (FWHM) of approximately 6.0, 6.0 and 9.8 \AA\,for the SB1, SB2, and SB3 bundles, respectively\footnote{These values correspond to the old camera mounted on WYFFOS, that was used for the present observations. From August 2004 a new camera was commissioned for the instrument. See more details in http://www.iac.es/proyecto/integral}. The covered spectral range of interest was $\lambda\lambda$4500-7000 \AA\,rest-frame. 
The reduction and calibration of the IFS data were performed inside the IRAF\footnote{The IRAF software is distributed by the National Optical Astronomy Observatory (NOAO), which is operated by the Association of Universities for Research in Astronomy (AURA), Inc., in cooperation with the National Science Foundation.} environment, and followed the standard procedures applied to this type of data (see Arribas et al.\,1997 and references therein). 
Full details of the observations and data reduction can be found in Garc\'{\i}a-Mar\'{\i}n et al.\,(2009).

Complementary \textit{HST} archive images were also used. Specifically, we used optical WFPC2 F439W, F606W, and F814W data, available for 12, 12, and 100\% of the ULIRGs under study, and near-IR NICMOS F160W data, available for 60\% of the sample. Filters F439W, F814W, and F160W, are the \textit{HST} analogs to the ground-based Johnson-Cousins B, I, and H, respectively (Origlia \& Leitherer 2000). The filter F606W is equivalent to a wide V- filter. In addition, F435W ACS images, equivalent to the Johnson-B filter and available for 29\% of the sample, were also used. The \textit{HST} images were calibrated on the fly, with the best available reference files. In the case of NICMOS data, the combination of the dithered individual exposures was repeated. All images were sky (or background) subtracted.

   \begin{figure*}[t]
   \begin{minipage}{\textwidth}
   \centering
   \includegraphics[width=0.8\textwidth]{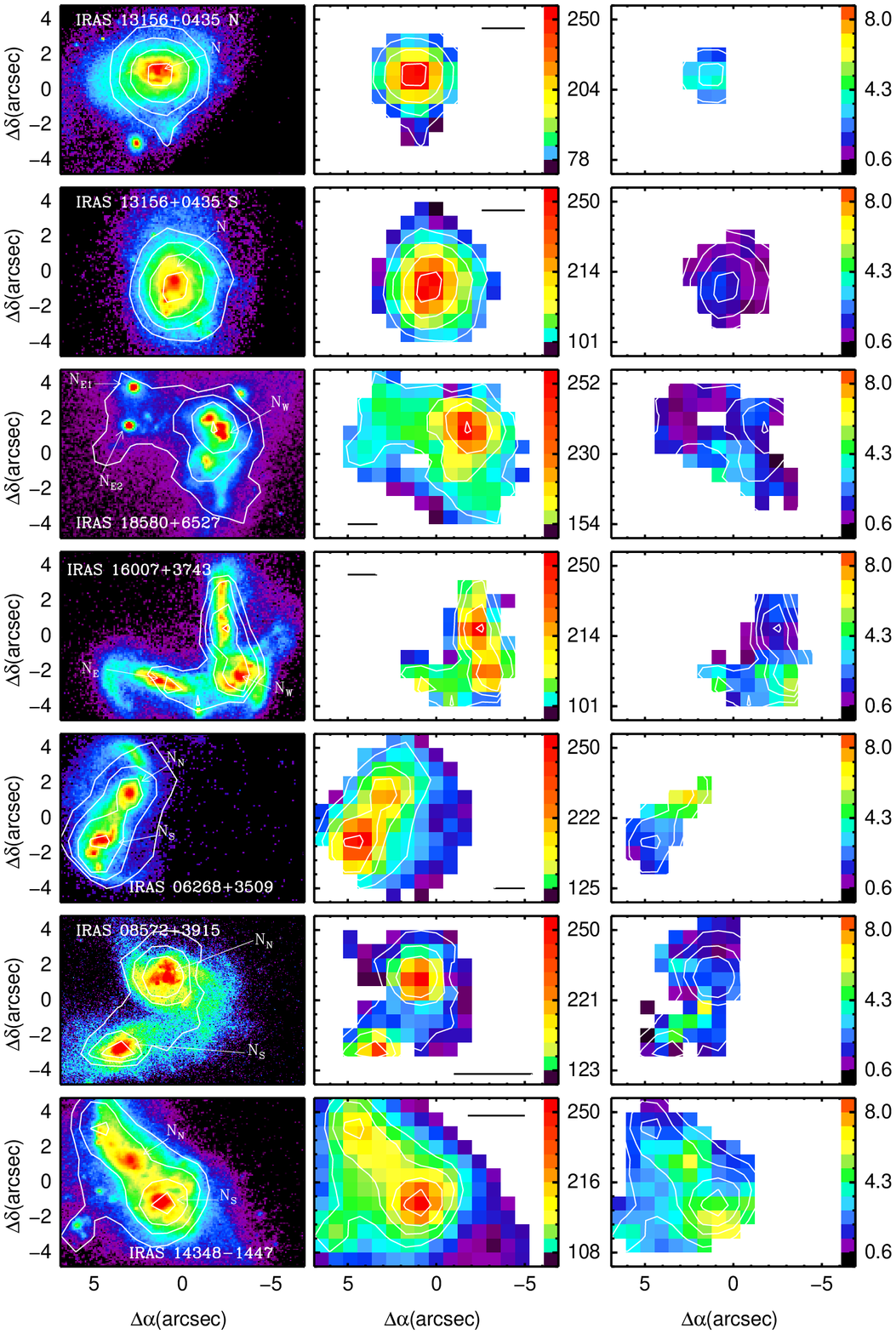}
      \caption{\textbf{a}. From left to right, HST WFPC2/F814W image, H$\alpha$ and extinction $A_{\rm{V}}$ INTEGRAL maps for a set of ULIRGs. The extinction maps were derived using the H$\alpha$/H$\beta$ line ratio. The HST and H$\alpha$ maps are represented in logarithmic scale, whereas for the extinction map the scale is linear. As a reference, the white contours shown on every map represent different H$\alpha$ intensity levels of each galaxy. For the H$\alpha$ map, we included a color code, which is given in relative flux units. Each pixel of the INTEGRAL maps represents the fiber size of the fiber used (0\farcs9). The horizontal scale represents 5~kpc. Galaxies orientation is north up, east to the left.}
   \label{Extincion1}
   \end{minipage}
   \end{figure*}

\addtocounter{figure}{-1}
   \begin{figure*}[t]
   \begin{minipage}{\textwidth}
   \centering
   \includegraphics[width=0.75\textwidth]{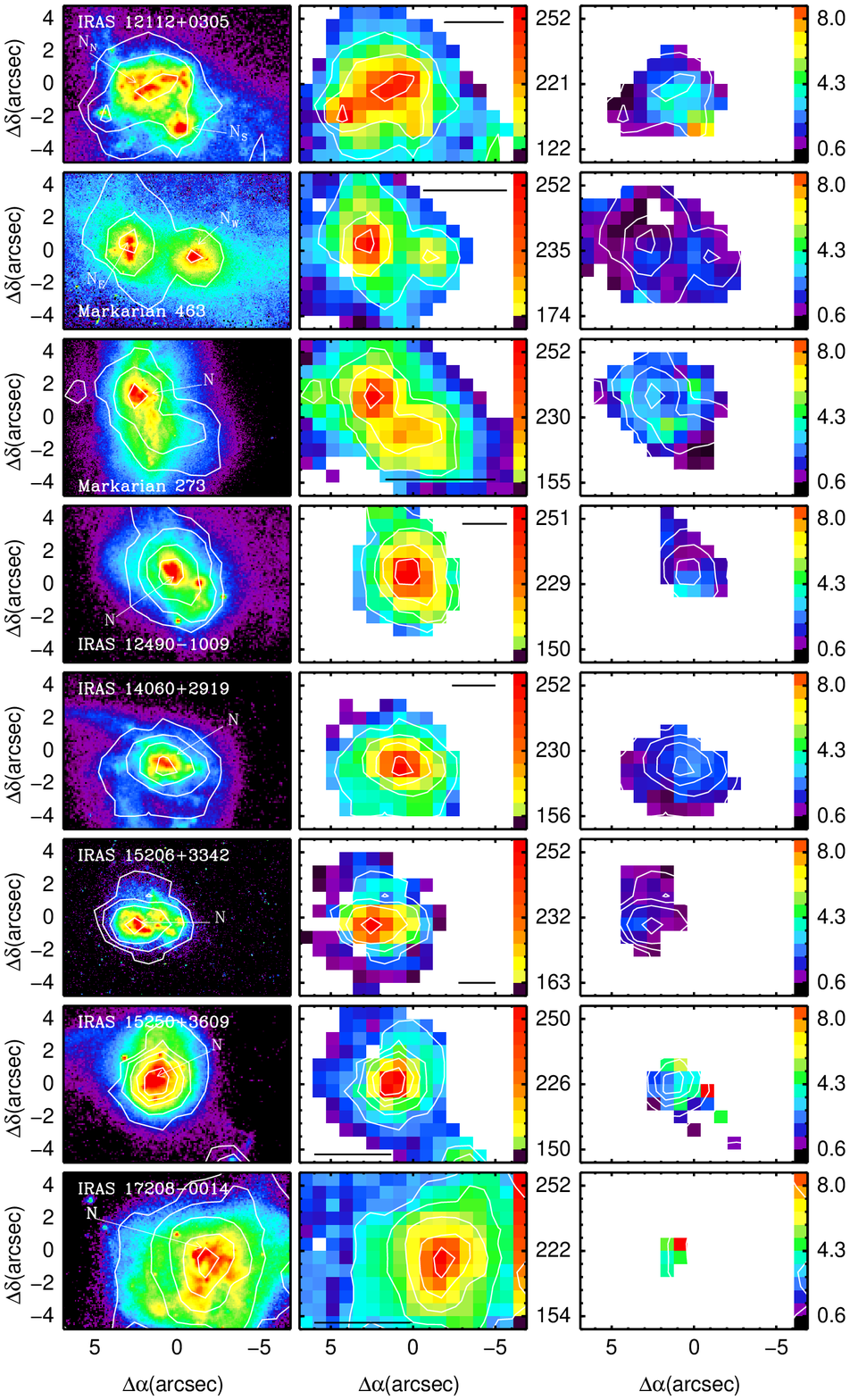}
      \caption{\textbf{b}. Same as Fig.\,\ref{Extincion1}\textbf{a}, but for a different set of ULIRGs.}
   \label{Extincion2}
   \end{minipage}
   \end{figure*}

\addtocounter{figure}{-1}
   \begin{figure*}[t]
   \begin{minipage}{\textwidth}
   \centering
   \includegraphics[width=0.6\textwidth]{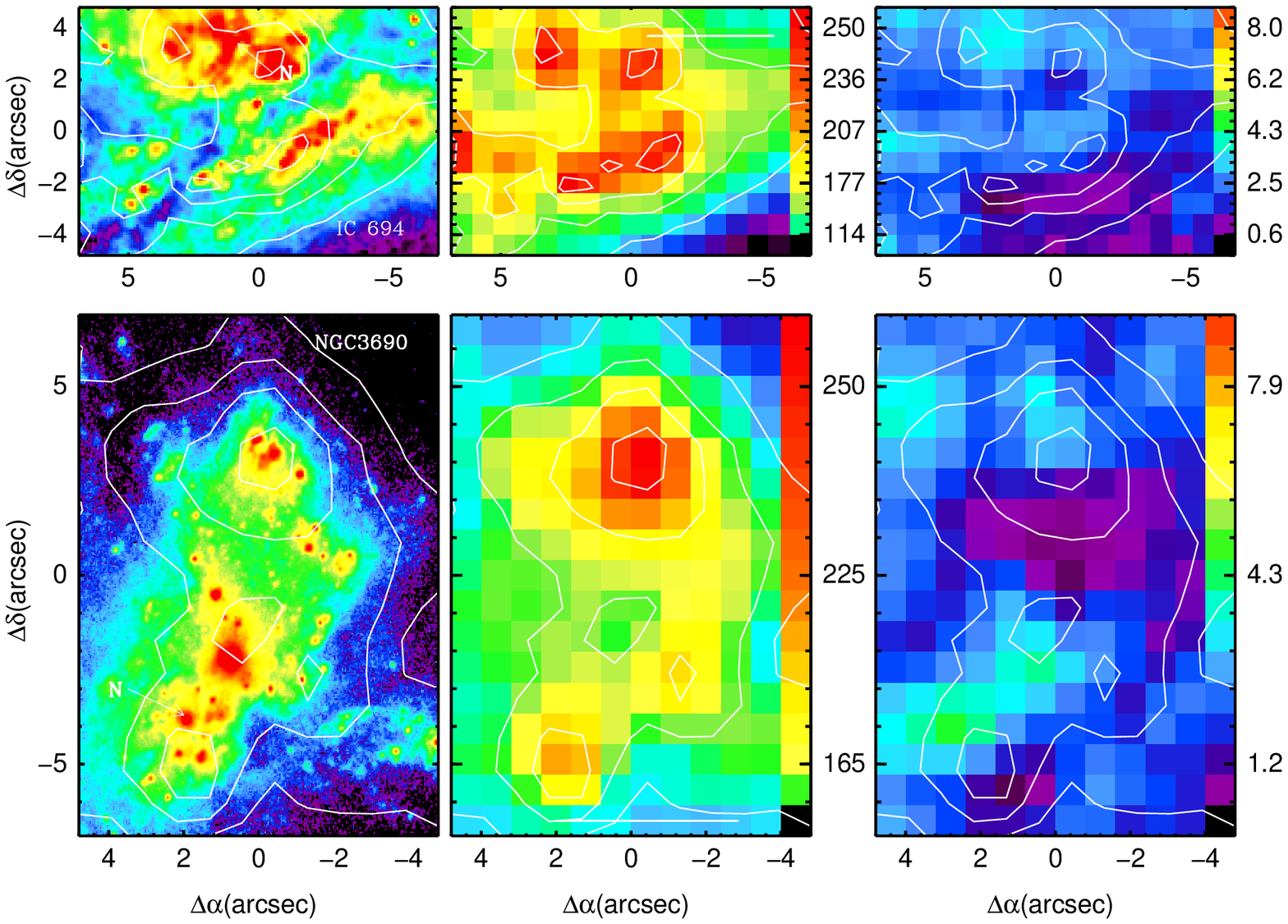}
      \caption{\textbf{c}. Same as Fig.\,\ref{Extincion1}\textbf{a}, but for Arp~299, an interacting galaxy formed by NGC~3690 (West) and IC~694 (East). In this case the scale represents 1~kpc.}
   \label{Extincion3}
   \end{minipage}
   \end{figure*}
\addtocounter{figure}{-1}
   \begin{figure*}[t]
   \begin{minipage}{\textwidth}
   \centering
   \includegraphics[width=0.7\textwidth]{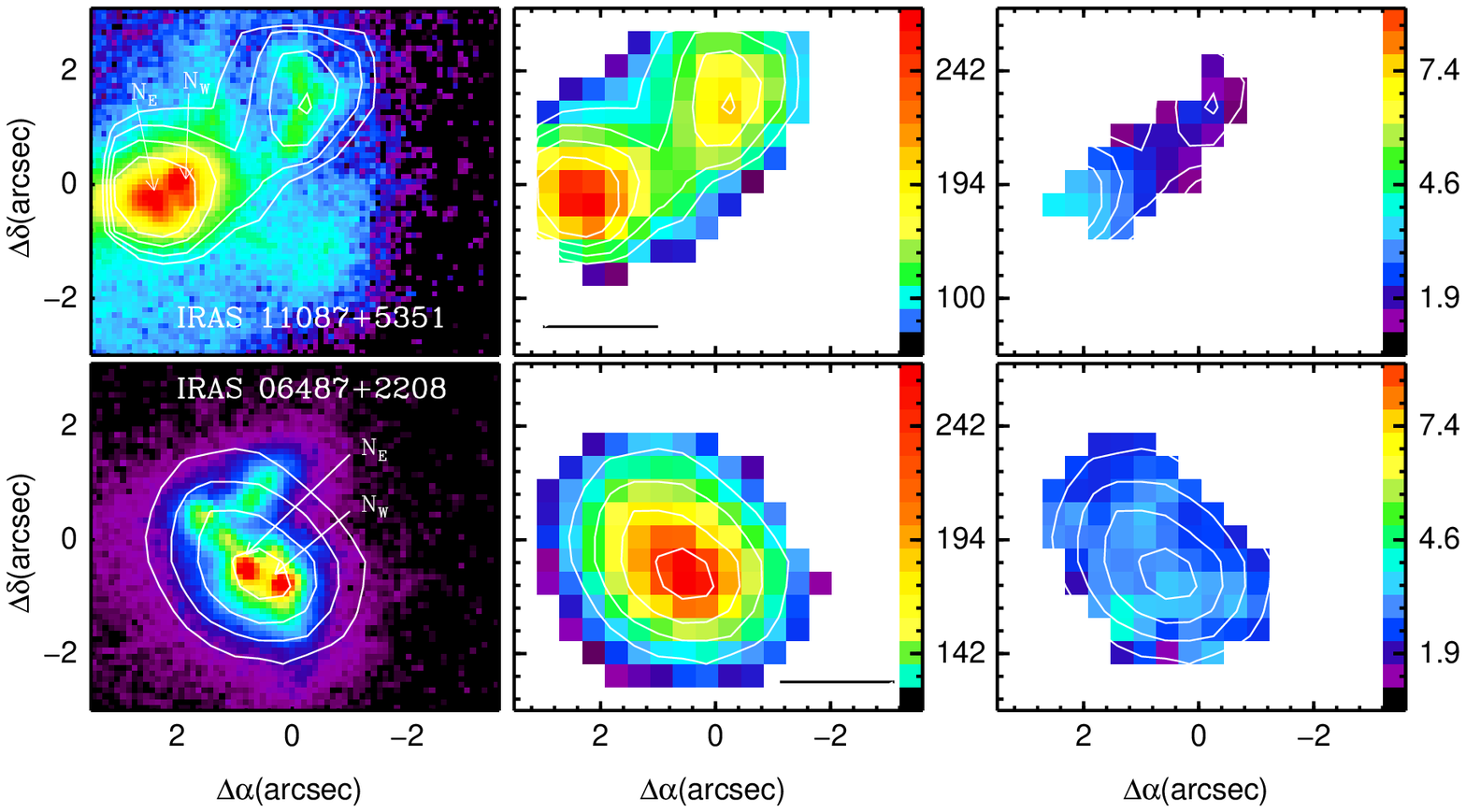}
      \caption{\textbf{d}. Same as Fig.\,\ref{Extincion1}\textbf{a}, but for two different ULIRGs observed with a fiber size of 0\farcs45.}
   \label{Extincion4}
   \end{minipage}
   \end{figure*}

   \begin{figure*}[t]
   \begin{minipage}{\textwidth}
   \centering
   \includegraphics[width=0.3\textwidth]{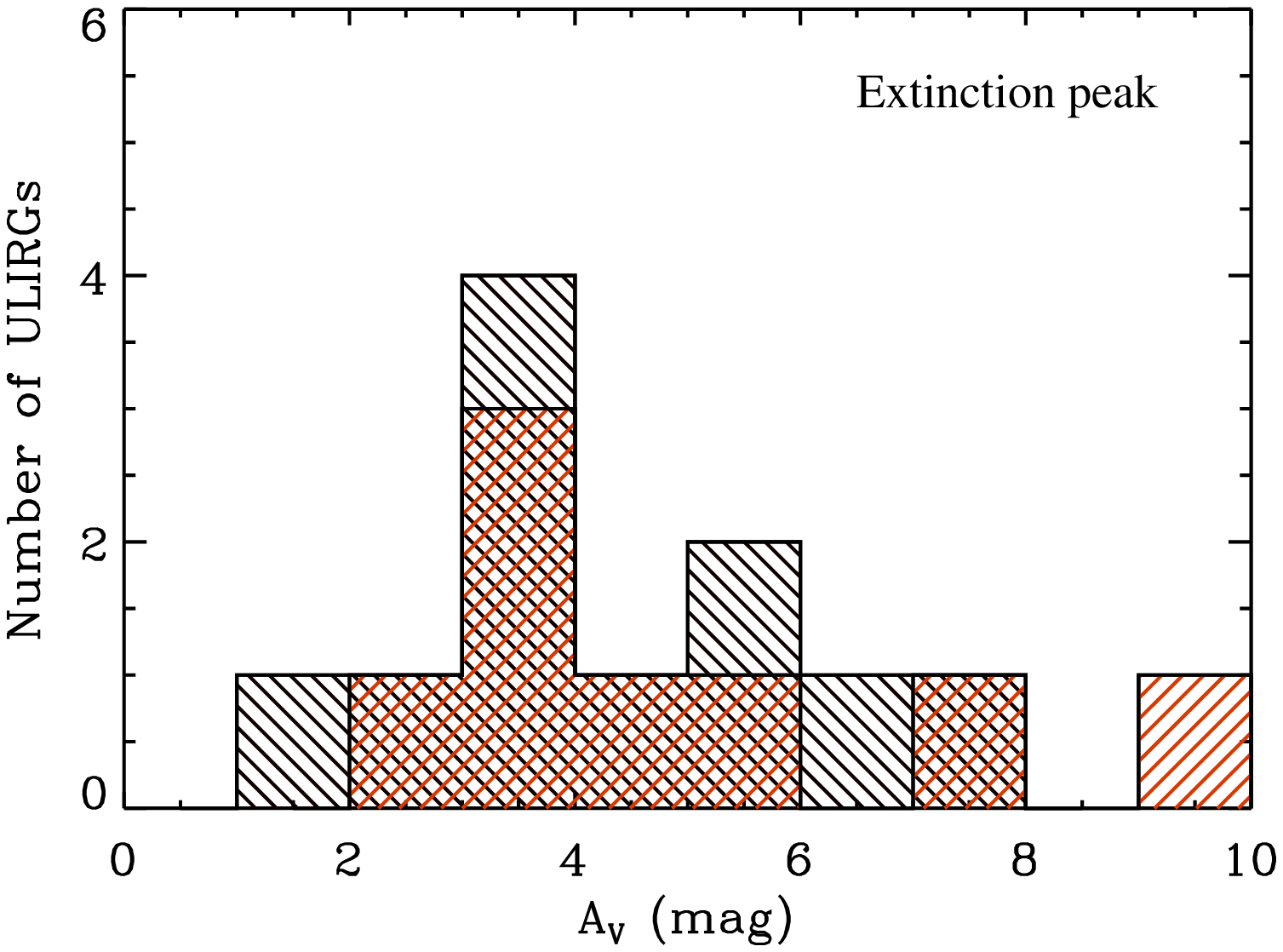}
   \includegraphics[width=0.3\textwidth]{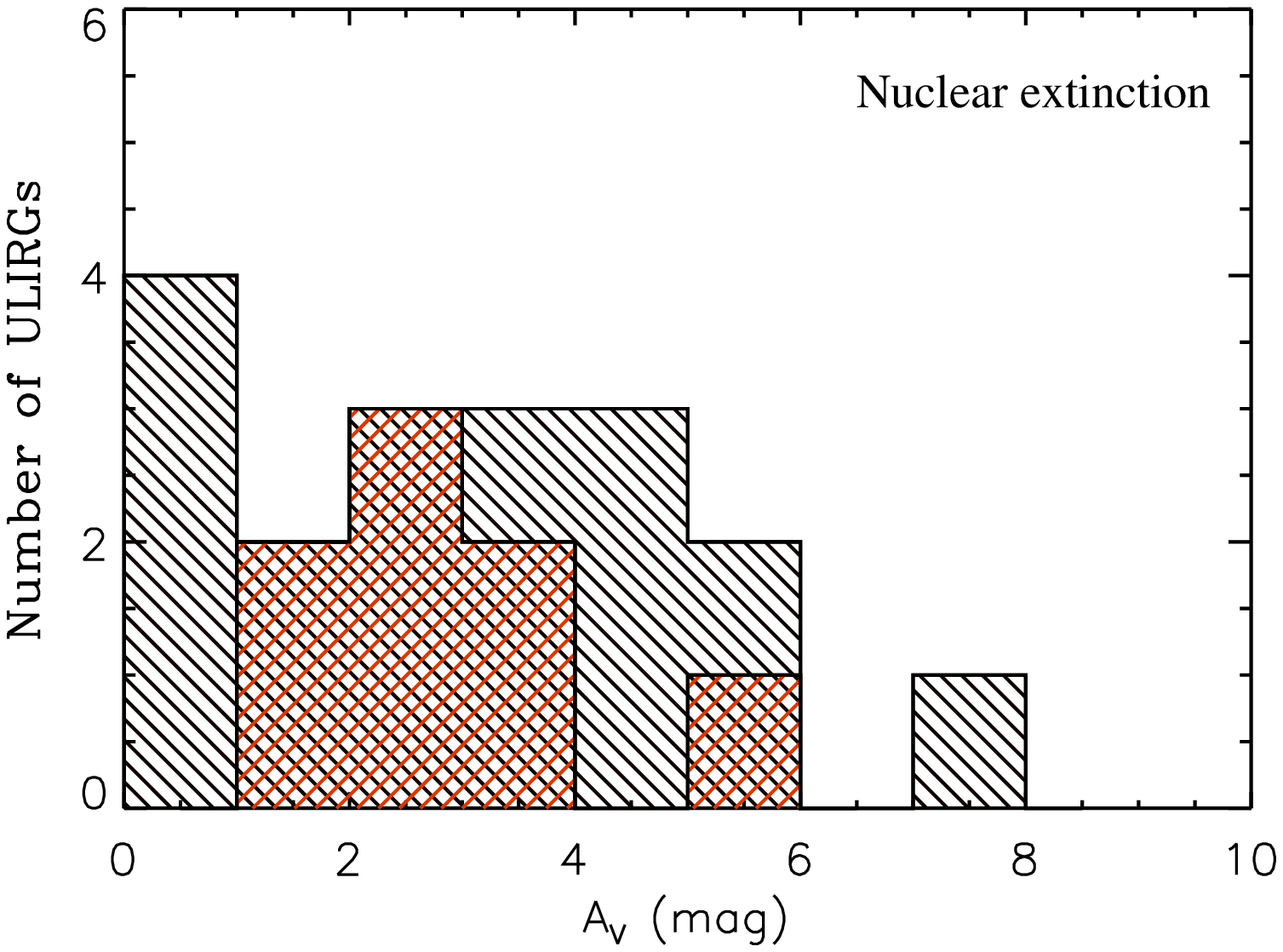}
   \includegraphics[width=0.3\textwidth]{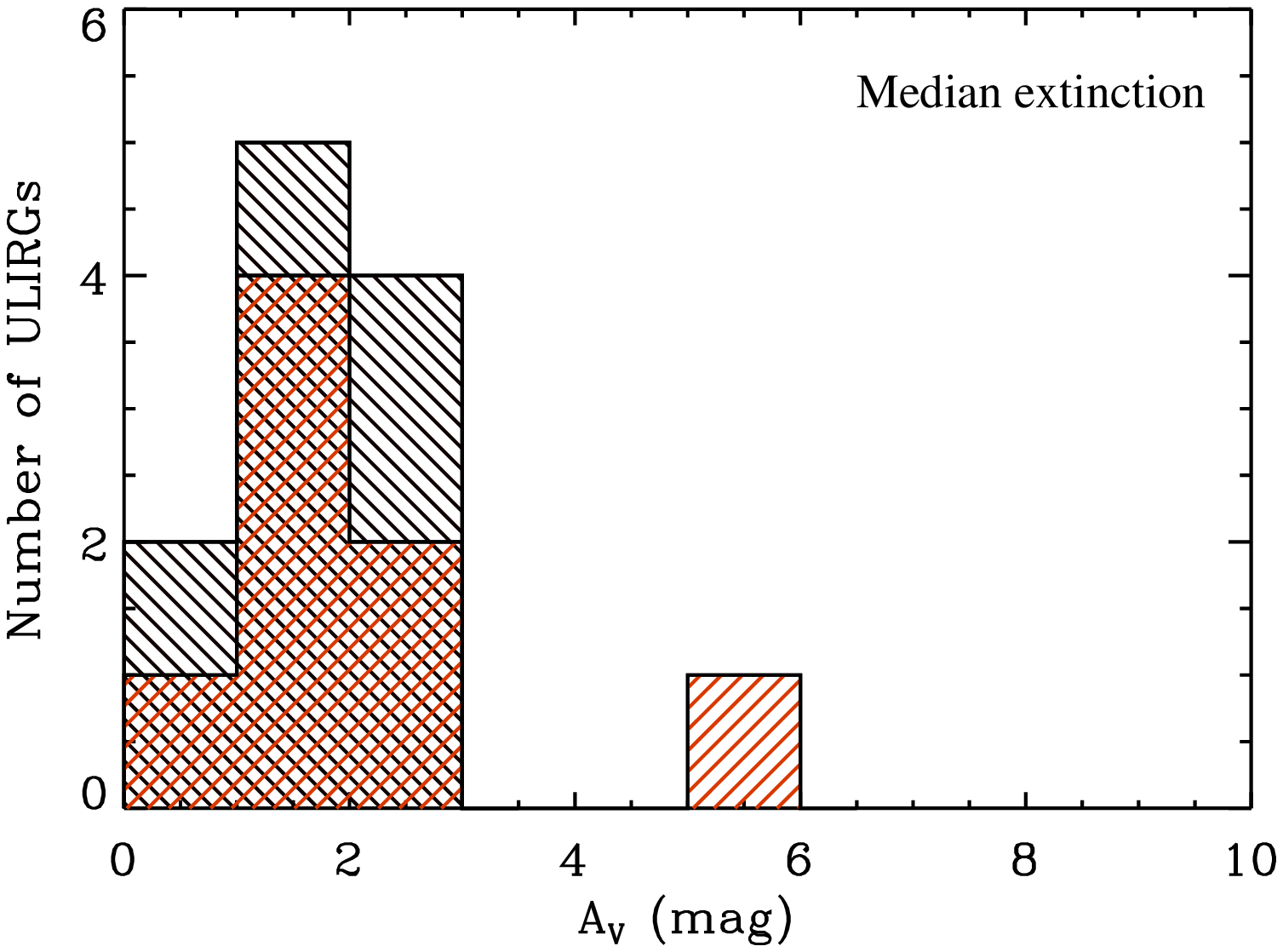}
      \caption{Distribution of the extinction values in the sample of ULIRGs. The black dashed histogram indicates pre-coalescence systems. The red dashed histogram indicates post-coalescence galaxies. \textbf{Left panel:} Distribution of the sample of galaxies according to the peak of extinction for each ULIRG. \textbf{Center panel:} Galaxy distribution according to the extinction in the nuclear regions. \textbf{Right panel:} Distribution of the sample of galaxies according to the median extinction of the systems. Note that in the cases of IRAS~13156+0435 and Arp~299, we derived the median for each galaxy of the interacting system using the individual pointings with the SB2 bundle.}
   \label{HistExtincion}
   \end{minipage}
   \end{figure*}

\section{Data analysis}
In all galaxies with available information, the two-dimensional structure of the  extinction/dust was derived using the H$\alpha$/H$\beta$ line ratio (Balmer decrement), a foreground dust screen model and a mean interstellar extinction law based on Savage \& Mathis (1979). The line fluxes were obtained by fitting each emission line to a single Gaussian function using the DIPSO package (Howarth \& Murray 1988) inside the STARLINK environment\footnote{See http://www.starlink.rl.ac.uk/.}. The presence of underlying stellar hydrogen absorption lines was not detected and therefore no correction was applied when measuring the flux of the corresponding emission lines. This would introduce a small overestimate in the values derived for the optical extinction since the equivalent widths of the hydrogen emission lines are larger than 40 \AA\ in the extended high surface brightness regions where they are detected, much larger than the
{{$EW_{\rm{abs}}$}(H$\alpha$)=$EW_{\rm{abs}}$}(H$\beta$)$\sim$2 \AA\ of the underlying stars.

It is known that different emission line ratios infer different extinction values, indicating that the observed recombination lines originate at different depths in these dusty regions. For this reason, especially in dust-enshrouded environments such as the central regions of ULIRGs where large quantities of dust are detected, the Balmer decrement provides only a lower limit to the true extinction. A clear example of this is the nucleus of the galaxy IC~694, a member of the pre-coalescence system Arp~299. The nuclear extinction derived using the Balmer decrement is {{$A_{\rm{V}}$}}$\simeq$3.0, whereas with the Pa$\alpha$/H$\alpha$ line ratio it is {{$A_{\rm{V}}$}}$\simeq$6.0 (Garc\'{\i}a-Mar\'{\i}n et al.\,2006).   

The \textit{HST} ACS, WFPC2, and NICMOS images were used to obtain the optical (B or V and I bands) and near-IR (H band) effective radius for each of the sample galaxies, defined as the radius that encloses half the
luminosity of the galaxy. It was derived using sky- or
background-corrected images, using apertures of increasingly large radius being centered on the optical nucleus of each galaxy. For the pre-coalescence systems, contamination from neighboring galaxy was prevented by the use of masks. In all cases, we include the outer low surface brightness regions of the galaxies, until the signal was at the same level as the background.

\section{Results and discussion}

\subsection{Two-dimensional H$\alpha$/H$\beta$-based extinction in pre- and post-coalescence systems}

The two-dimensional H$\alpha$/H$\beta$-based extinction maps of the sample galaxies cover areas of several kpc on a side, ranging from about 2.5 kpc in highly obscured systems (e.g., IRAS 17208$-$0014) to 10-12 kpc in galaxy pairs 
(e.g., IRAS 08572$+$3915 or IRAS 14348$-$1447). The angular resolution of the data provides extinction values on scales of between 0.7 and 3 kpc, depending on the redshift of the source (see specific values for each galaxy in Table\,\ref{Table1}).
On these scales, the extinction maps (Fig. \ref{Extincion1}) show that the dust in ULIRGs is not uniformly distributed. It shows instead a patchy structure including almost transparent regions with very low dust content, and other regions that are highly absorbed where, in the optical domain, only a lower limit to the extinction can be derived. 

The visual extinction ({$A_{\rm{V}}$}=3.1$\times$E(B-V)) ranges from {$A_{\rm{V}}$}$\leq$0.2 mag, generally measured in 
the external regions, to {$A_{\rm{V}}$}$\simeq$9.0 mag, measured in the central kpc of of IRAS 17208-0014. The highest extinctions are usually being measured in the nuclear regions, whose values cover a broad range of {$A_{\rm{V}}$} between 0.6 and 7.7 mag, 69\% of the nuclei having {$A_{\rm{V}}$}$>$2.0 mag.
These high extinction values are similar to those measured in the Pa$\alpha$ emitting regions of LIRGs, whose average extinctions are found to be {$A_{\rm{V}}$}=3 -- 6~mag (Alonso-Herrero et al.\,2006) but lower than the values derived for ULIRGs using the near-IR Pa$\alpha$/Br$\gamma$ line ratio, where visual extinctions ({$A_{\rm{V}}$}) in excess of 10 magnitudes are relatively common in ULIRGs (Murphy et al. 2001). Both optical and near-IR extinction values are also systematically lower than the Spitzer 
mid-IR based ones derived for local ULIRGs, which are found to be $<$20 -- 30 mag for PAH-emitting normal starbursts, and $>>$30 mag for centrally concentrated energy sources with no PAH emission (Imanishi et al.\,2007). 

Pre- and post-coalescence systems are in different phases of the evolutionary merging process, and therefore their dust 
and gas content and distribution could reflect this. Pre- and post-coalescence systems do have similar nuclear extinctions with average values of 3.0 and 2.5 mag, respectively. However, while while extinction values for post-coalescence nuclei tend to be between 2 and 3 mag of visual extinction, the values for pre-coalescence nuclei appear to be more widely distributed over the entire 0.5 mag $\leq$ {$A_{\rm{V}}$} $\leq$ 8 mag range (Fig.\,\ref{HistExtincion} center and Table 1 for specific values). 

The median extinction derived for entire galaxies (Fig.\,\ref{HistExtincion} right) show a clear indication that
extinction in the circumnuclear and external regions is substantially lower than in the nuclei. While only 31\% of all nuclei have visual extinctions of 2 mag or less, 60\% of the galaxies do have median extinctions with those low values.
Of all the systems, pre-coalescence ULIRGs show on average a difference of 1.2 mag between the median visual extinction for the entire galaxy (1.8 mag) and the nuclear regions (3.0 mag). Post-coalescence systems do however show similar values (2.1 and 2.5 mag for median and nuclear extinctions, respectively). This behavior is clearly explained as a consequence of the different morphology of 
pre- and post-coalescence systems. Pre-coalescence systems contain several low-extinction H$\alpha$ high-surface 
brightness regions at distances of several kpc away from the nucleus associated with star formation along the tidal 
tails and candidates to tidal dwarf galaxies (Monreal-Ibero et al. 2006). Post-coalescence systems are on the other hand
more compact and dominated by the emission from the nucleus. The importance of these effects in deriving star formation rates in dust-enshrouded systems is discussed next.

\subsection{Implications of the extinction structure in the derivation of the Star Formation Rates}

Star formation rates (\textit{SFR}) in galaxies are derived using different tracers, from ultraviolet light and optical forbidden
lines ([O\,{\sc ii}] 3727\AA), to hydrogen recombination lines (e.g. H$\alpha$, Pa$\alpha$, Br$\gamma$), and IR broad-band flux (see e.g. Kennicutt 1998). The H$\alpha$ line is the most intense emission line in the optical, and shifts into the near-
and mid-IR
spectral range for intermediate and high$-z$ galaxies. Because of that, it is one of the most common spectral features used
in deriving \textit{SFR} in all kinds of galaxies at all redshifts. 

For obscured and dust-enshrouded galaxies such the low$-z$ ULIRGs and the high$-z$ submillimeter and 24 
$\mu$m bright Spitzer galaxies, the extinction correction applied has important consequences in the reliability of the H$\alpha$-based \textit{SFR}. The \textit{SFR} and H$\alpha$ luminosity are directly proportional as given by the expression

\begin{equation}
SFR(M_{\odot}\,yr^{-1})= 7.9 \times 10^{-42} L_{\rm{H}\alpha} (erg\,s^{-1})
\end{equation}

\noindent
for a continuous star formation, a Salpeter initial mass function with upper and lower mass limits of 100 and 0.1 M$_{\odot}$, respectively, and assuming that all potentially ionizing photons ionize the surrounding interstellar medium (Kennicutt 1998). Most dust extinction in low$-z$ ULIRGs to date are obtained through narrow (typically 1 to 1.5 arcsec) long-slit spectra along a given orientation centered on the nucleus of the galaxy (if post-coalescence), or connecting the nuclei of interacting pairs (if pre-coalescence). Observational  difficulties arising from differential atmospheric refraction and positioning of the galaxy on the slit translate into an increased uncertainty in the measured values. Since the width of the slit does not cover the entire H$\alpha$ emitting regions (see Figs. 1-4 for angular extent and structures), the extinction values derived from these data will be dominated by the emission from the nucleus. This would 
underestimate the H$\alpha$ luminosity, and therefore the \textit{SFR}, by an average factor of about 4 and 7 in the pre- and post-coalescence ULIRGs, respectively (see Table 2 for the specific ratios of the observed nuclear to total H$\alpha$ flux (Col. 3), and the corresponding extinction corrected luminosities (Cols. 5 and 8) for each galaxy).

For the typical median ULIRG extinction ({$A_{V}$}$\simeq$2.0 see Fig.\,\ref{HistExtincion} right panel), the use of observed H$\alpha$ luminosities underestimates the \textit{SFR} by a factor 6 with respect to the extinction-corrected value.
Only 28\% of the galaxies in the sample have de-reddened \textit{SFR} of less than 20 M$\odot$ yr$^{-1}$, whereas for the uncorrected \textit{SFR} this percentage increases to 72\%. The total de-reddened \textit{SFR} of the present ULIRGs sample ranges from about 10 to 
300 M$\odot$ yr$^{-1}$, for a similar range of \textit{SFR} values for pre- and post-coalescence systems. 
Finally, we note that because of the lower S$/$N ratio of the H$\beta$ line, the extinction maps do not cover the same field as the  H$\alpha$ maps (see from Figs.\,1 to 4). This would lead to an additional underestimate of the two-dimensional extinction corrected H$\alpha$ fluxes (and thus of the \textit{SFR} values). In any case, and since this mainly affects the outer low-surface brightness regions, no significant changes should be expected.

The \textit{SFR} is also found to be proportional to the {$L_{\rm{IR}}$}. Considering continuous bursts of 10-100 Myr and applying the 
same range of IMF parameters as used in Eq. (1) above, the proportionality is given by Kennicutt (1998)

\begin{equation}
SFR(M_{\odot}\,yr^{-1})= 4.5 \times 10^{-44} L_{\rm{IR}} (erg\,s^{-1}).
\end{equation}

\noindent The \textit{SFR} as derived from the IR luminosity is systematically higher than that obtained from the H$\alpha$ measurements (see Table 3). The ratio {$SFR(IR)/SFR(L(H\alpha$)$_{\rm{corr}})$} covers a wide range, from about 2 to 10, and tends to be smaller in galaxies with higher nuclear extinctions, i.e., IRAS12112+0305, IRAS14348-1447, and IRAS16007+3743. This is because of the large amounts of dust obscuring the nuclear regions, which cause the innermost regions to be more optically thick to optical than far infrared radiation. Thus, even by applying the two-dimensional extinction corrections to the 
observed H$\alpha$ fluxes, the derived \textit{SFRs} would represent 10\% to 50\% of the IR-based \textit{SFR} value, which is assumed to be free of extinction effects, and therefore represent the true \textit{SFRs}.

\begin{table*}
\begin{minipage}[t]{\textwidth}
\centering
\renewcommand{\footnoterule}{}  
\caption{Observed H$\alpha$ fluxes and extinction corrected luminosities for the sample galaxies.}   
\label{Table2}      
\tabcolsep0.1cm
\begin{tabular}{lcrccccc}       
\noalign{\smallskip}
\hline
\noalign{\smallskip}
Galaxy & {$F(\rm{H}\alpha)_{\rm{obs}}$}\footnote{Integrated H$\alpha$ observed flux, with units $\times$10$^{-14}$ erg s$^{-1}$ cm$^{-2}$. Note that in the cases of IRAS~13156+0435 and Arp~299 we derived the integrated flux for each galaxy of the interacting pair   using their individual pointings.} &  {$F(\rm{H}\alpha)_{\rm{nuclear/total}}$}\footnote{Ratio between the H$\alpha$ nuclear, identified as the peak position in the optical continuum and total observed fluxes.} & {$L(\rm{H}\alpha)_{\rm{obs}}$}\footnote{Observed H$\alpha$ luminosity derived from the integrated flux in column 2. Units in $\times$10$^{41}$ erg s$^{-1}$} & {$L(\rm{H}\alpha)_{\rm{ext corr}}$}\footnote{Integrated two-dimensional extinction corrected H$\alpha$ luminosity, with units in $\times$10$^{41}$ erg s$^{-1}$} & {$L(\rm{H}\alpha)_{\rm{med-ext}}$}\footnote{Integrated H$\alpha$ extinction-corrected luminosity, derived using the total observed flux and the median extinction, with units in $\times$10$^{41}$ erg s$^{-1}$} &{$L_{\rm{nuc}}(\rm{H}\alpha)_{\rm{obs}}$}\footnote{Observed nuclear H$\alpha$ luminosity with units in $\times$10$^{41}$ erg s$^{-1}$} & {$L_{\rm{nuc}}(\rm{H}\alpha)_{\rm{corr}}$}\footnote{Extinction corrected nuclear H$\alpha$ luminosity, derived using the nuclear extinction of each individual galaxy. Units in $\times$10$^{41}$ erg s$^{-1}$} \\ 

\hline
IRAS~13156+0435N   & 0.9   & 0.12     & 3.0   & 21.0   &  27.1  & 0.36  & 4.6     \\
IRAS~13156+0435S   & 1.2   & 0.09     & 3.9   & 9.0    &  6.1   & 0.35  & 1.3     \\
IRAS~18580+6527    & 2.5   & [E] 0.01 & 21.6  & 78.0   &  68.2  & 0.22  & 0.4     \\
                   &       & [W] 0.28 &       &        &        & 6.05  & 15.9    \\
IRAS~16007+3743    & 2.6   & [E] 0.03 & 25.1  & 152.0  &  134.7 & 0.75  & 17.7    \\
                   &       & [W] 0.22 &       &        &        & 5.52  & 97.0    \\
IRAS~06268+3509    & 0.6   & [N] 0.09 & 4.7   & 121.0  &  28.0  & 0.42  & 66.5    \\
                   &       & [S] 0.18 &       &        &        & 0.85  & 4.4     \\
IRAS~08572+3915    & 5.8   & [N] 0.08 & 4.6   & 15.5   &  21.1  & 0.37  & 2.0     \\
                   &       & [S] 0.05 &       &        &        & 0.23  & 0.3     \\
IRAS~14348$-$1447  & 15.0  & [N] 0.03 & 25.6  & 381.0  &  333.0 & 0.77  & 52.3    \\
                   &       & [S] 0.08 &       &        &        & 2.05  & 31.5    \\
Mrk~463            & 36.9  & [E] 0.25 & 21.8  & 56.0   &  48.2  & 5.45  & 10.4    \\
                   &       & [W] 0.02 &       &        &        & 0.44  & 1.8     \\
Arp~299/NGC~3690   & 181.0 &     0.03\footnote{For this particular galaxy we have calculated the ratio using the position of the true nucleus of the galaxy, which does not coincide with the optical one (Garc\'{\i}a-Mar\'{\i}n et al 2006).} & 4.0   & 29.0   &  21.5  & 0.12  & 0.9     \\
Arp~299/IC~694     & 89.0  &     0.03$^{h}$ & 2.0   & 16.0   &  11.9  & 0.06  & 6.2     \\
IRAS~12112+0305    & 5.8   & [N] 0.08 & 7.6   & 213.0  &  34.0  & 0.60  & 9.0     \\
                   &       & [S] 0.02 &       &        &        & 0.15  & 155.0   \\
IRAS~06487+2208    & 2.5   &     0.19 & 13.9  & 117.0  &  96.3  & 2.64  & 23.3    \\
IRAS~11087+5351    & 0.8   &     0.12 & 4.4   & 11.0   &  12.0  & 0.53  & 7.4     \\
Mrk~273\footnote{The data for Mrk~273 do not have an absolute flux calibration. Hence, only the relative measurement between the nuclear and integrated H$\alpha$ flux is given here.} &    ---   &     0.08 &  ---     &   ---     &   ---     &   ---    &   ---      \\
IRAS~12490$-$1009  & 1.0   &     0.13 & 2.6   & 12.0   &  8.1   & 0.34  & 0.9     \\
IRAS~14060+2919    & 2.3   &     0.18 & 8.2   & 45.0   &  30.8  & 1.48  & 10.7    \\
IRAS~15206+3342    & 42.7  &     0.38 & 171.8 & 322.0  &  348.7 & 65.3  & 158.5   \\
IRAS~15250+3609    & 4.4   &     0.26 & 3.2   & 87.5   &  25.0  & 0.83  & 4.4     \\
IRAS~17208$-$0014  & 17.9  &0.10      & 7.7 & 160.2 & ---\,\footnote{As shown in Fig.\,1b, the area of this galaxy with available extinction values is limited, also presenting a vey large peak value. The use of this value for correcting the integrated flux is not valid and therefore is not included in the table.} & 0.7 & 43.0 \\
\hline                                   
\end{tabular}
\end{minipage}
\end{table*} 


\begin{table*}
\begin{minipage}[t]{\textwidth}
\centering
\renewcommand{\footnoterule}{}  
\caption{H$\alpha$- and IR-based star formation rates for the sample of ULIRGs.}   
\label{Table3}      
\tabcolsep0.1cm
\begin{tabular}{lccccc}       
\noalign{\smallskip}
\hline
\noalign{\smallskip}
\hline
\noalign{\smallskip}
Galaxy & {$log(L_{\rm{IR}}$)}\footnote{IR luminosity derived following Sanders \& Mirabel (1996)} & {$SFR(L(\rm{H}\alpha)_{\rm{obs}}$)}\footnote{SFR derived using the total observed H$\alpha$ flux given in column 2 of Table 2. Units are M$\odot$ yr$^{-1}$} & {$SFR(L(\rm{H}\alpha)_{\rm{med-ext}}$}\footnote{SFR derived using the H$\alpha$ median extinction corrected values as specified in Table 2 column 6. Units are M$\odot$ yr$^{-1}$}) & {$SFR(L(\rm{H}\alpha)_{\rm{corr}}$)}\footnote{SFR calculated using the two-dimensional extinction corrected H$\alpha$ luminosity as given in Table 2 column 5. Units are M$\odot$ yr$^{-1}$}& {$SFR(IR)$}\footnote{SFR based on the IR-based luminosity values as given in column 2 of the present table. Units are M$\odot$ yr$^{-1}$.}\\ 

\hline
IRAS~13156+0435   & 12.13 & 5.5  & 26.2  & 23.7  & 232.9 \\
IRAS~18580+6527\,\footnote{Strong evidence for a Seyfert 2 nucleus, for which we have not corrected this values}   & 12.26 & 17.1& 53.8  & 61.6  & 318.7 \\
IRAS~16007+3743   & 12.11 & 19.8 & 106.4 & 120.  & 222.5 \\
IRAS~06268+3509   & 12.51 & 3.7  & 22.1  & 95.6  & 493.5 \\
IRAS~08572+3915   & 12.17 & 3.6  & 16.7  & 12.2  & 255.5 \\
IRAS~14348-1447   & 12.39 & 20.2 & 263.1 & 301.0 & 424.1 \\
Mrk~463           & 11.81 & 17.2 & 38.0  & 44.2  & 111.5 \\
Arp~299           & 11.81 & 4.7  & 26.3  & 35.3  & 111.5 \\
IRAS~12112+0305   & 12.37 & 6.0  & 26.9  & 168.3 & 405.0 \\
IRAS~06487+2208   & 12.57 & 11.0 & 76.0  & 92.4  & 641.8 \\
IRAS~11087+5351   & 12.13 & 3.5  & 9.5   & 8.7   & 233.0 \\
Mrk~273$^{f}$     & 12.18 & ---     &---       &  ---     & 216.5 \\
IRAS~12490-1009   & 12.07 & 2.0  & 6.4   & 9.5   & 203.0 \\
IRAS~14060+2919   & 12.18 & 6.4  & 24.3  & 35.5  & 261.5 \\
IRAS~15206+3342   & 12.27 & 135.7& 275.5 & 254.3 & 321.6 \\
IRAS~15250+3609   & 12.09 & 2.5  & 19.7  & 69.1  & 212.5 \\
IRAS~17208-0014   & 12.43 & 6.1  & ---   & 126.5 & 465.0 \\
\hline                                   
\end{tabular}
\end{minipage}
\end{table*} 



\subsection{Implications of the extinction structure for the derivation of the dynamical mass}

Dynamical masses ({$\propto \sigma^2 \times R_{\rm{hm}}$)} of galaxies are often derived from measurements of the velocity dispersion of the stars or gas, and assuming the radius ({$R_{\rm{hm}}$)} where half of the mass is contained, is known. While the velocity dispersion can be directly obtained, to first order, from the full-width-half-maximum of the absorption/emission line profiles, the half-mass radius is not a direct observable and has to be inferred from the half-light radius. 

The most common lines used to derive velocity dispersions in low-{\it z} ULIRGs and their high-{\it z} analogs are in the optical (e.g., H$\beta$ and H$\alpha$) and near-IR (e.g., Br$\gamma$, CO stellar absorption bands) rest-frame, and are therefore affected by non-uniform extinction (see Sect. 5.1). The behavior of the extinction in ULIRGs with large values in the nuclear regions and an outward decreasing gradient in the circumnuclear and extranuclear regions
(see Sect. 5.1 and Fig. 1), produces an apparent stellar light distribution that differs from that of the true mass distribution. On the one hand, for a given 
rest-frame wavelength, the half-light radius will always be an upper limit to the true $R_{\rm{hm}}$ as the light from 
stars in the circum- and extranuclear regions are less obscured than those in the nuclear regions. 
On the other hand, the extinction has a strong wavelength dependence becoming higher towards bluer wavelengths. 
For a given galaxy, the half-light radius derived in the optical will 
always be systematically larger than those derived in the near-IR. This effect is observed in our sample galaxies where
for those galaxies with available $HST$ optical (0.8$\mu$m, WFPC2 F814W) and near-IR (1.6 $\mu$m, NICMOS2 F160W) images
(see Table 1), the optical half-light radius is on average 2.2 times the corresponding value in the near-IR, in some systems up to 5 times larger. 
Therefore, in obscured systems such as local and high-z ULIRGs, the use of optical (red) rest-frame light will produce an average overestimate of the dynamical mass by a factor of 2.2 merely because of extinction effects. This factor could be even
larger if effective radii are measured in the blue (B) or visual (V) rest-frames (see specific values in Table 1), as frequently performed for in high$-z$ galaxies.

In addition to these extinction effects, other effects such as stellar population gradients can also be important. Since young (10 Myr or less) massive stars and older (100 Myr to Gyrs) stellar populations have mass-to-light ratios that differ by up to two orders of magnitude in the optical and near-IR (e.g. Starburst99, Leitherer et al. 1999), their non-uniform spatial distribution is also relevant to evaluating the uncertainty in the measured half-light radius, and therefore in the dynamical mass. With only two filters available for most of the sample galaxies, a detailed analysis of these effects is beyond the scope of the present work.

\section{Summary}
This paper has presented the study of the two-dimensional extinction structure of a representative sample of 17 ULIRGs, 9 pre-coalescence (i.e., interacting pairs with projected nuclear separations of at least 1.5 kpc), and 8 post-coalescence
(either single nucleus or double nuclei with projected separations of less than 1.5 kpc). The extinction maps are based on the measurement of the H$\alpha$/H$\beta$ line ratio obtained from integral field spectroscopy with the instrument INTEGRAL on the William Herschel telescope. In comparison with more classical narrow long-slit spectroscopy, this study infers the two-dimensional extinction structure of ULIRGs on kpc scales over areas of a few to several kpc on a side. The main results are summarized below:

\begin{itemize}
\item In agreement with previous studies, the analysis of our data has detected a very complex and patchy extinction structure in ULIRGs
on scales of kpc, from basically transparent regions to others so deeply embedded in dust that only a lower limit to their extinction can be derived ({$A_{\rm{V}}$}$\simeq$0.0 mag to {$A_{\rm{V}}$$\simeq$8.0 mag}). Nuclear extinction values of the present galaxy sample cover a broad range in A$_{V}$ from 0.6 to 8 mag, 69\% of the nuclei having A$_{V}$$>$2.0 mag. Extinction in the external regions is substantially lower than in the nuclei with 60\% of the ULIRGs in the sample having a median A$_{V}$ for the entire galaxy of less than two magnitudes.

\item ULIRGs classified as pre- and post-coalescence exhibit significant differences in their extinction properties. While
post-coalescence nuclei tend to cluster around {$A_{\rm{V}}$} values of 2 to 3 mag, pre-coalescence nuclei appear more homogeneously distributed over the entire 0.5 mag $\leq$ {$A_{\rm{V}}$} $\leq$ 6 mag range. Pre-coalescence ULIRGs show on average a difference of 1.2 mag between the nuclear ({$A_{\rm{V}}$} of 3.0 mag) and median extinction for the entire system ({$A_{\rm{V}}$} of 1.8 mag), while post-coalescence systems do show similar values (2.1 and 2.5 mag for median and nuclear extinctions, respectively). This behavior is explained as a consequence of the different distribution of the star-forming regions in pre- and post-coalescence systems. Pre-coalescence systems contain several 
low-extinction H$\alpha$ high-surface brightness regions at distances of several kpc away from the nucleus associated 
with star formation along the tidal tails. Post-coalescence systems are more compact and dominated by the emission from 
the nuclear regions. 

\item  Considering the median extinctions measured in the sample galaxies ({$A_{\rm{V}}$$\simeq$2.0}), the \textit{SFR} values based on the de-reddened H$\alpha$ fluxes increase by a factor 6 with respect to those derived from the observed H$\alpha$ fluxes. The H$\alpha$-based extinction-corrected \textit{SFR} ranges from about 10 to 300 M$\odot$ yr$^{-1}$. The de-reddened \textit{SFR} is less than
20 M$\odot$ yr$^{-1}$ in 28\% of the galaxies, whereas for the observed \textit{SFR} this percentage increases up to 72\%. No significant differences in the \textit{SFR} have been found between the pre- and post-coalescence systems.   

\item In all galaxies, the IR-based \textit{SFR} is higher than the de-reddened H$\alpha$ value. The ratio ranges from 2 to 10, with the smaller differences in galaxies with higher extinction in their nuclear regions. Thus, assuming IR luminosities measure the true \textit{SFR} in these systems, \textit{SFRs} based on full two-dimensional extinction corrected H$\alpha$ luminosities recover between 10\% and 50\% of the true \textit{SFR}. 
 
\item Because of the high nuclear extinctions and the outward decreasing extinction gradient, the optical (I-band) 
half-light radius 
in the sample of ULIRGs is on average a factor 2.2 larger than the corresponding value in the near-IR (H-band), and in some systems even a factor 5 larger. Dynamical masses in ULIRGs and their high$-z$ analogs would be on average overestimated by a factor 2.2 due merely to extinction effects if the effective radius of the galaxy is derived directly from observed rest-frame optical images. This factor could be even larger if bluer broad-band filters (i.e., V and B bands) were used to measure the
effective radius as often the case with high$-z$ galaxies. Additional stellar population effects could also be relevant because of their different mass-to-light ratios, as should be investigated by future detailed studies.

\end{itemize}

\begin{acknowledgements}
This paper uses the plotting package \texttt{jmaplot}, developed by Jes\'us Ma\'{\i}z-Apell\'aniz. http:$//$dae45.\break iaa.csic.es:8080$/$$\sim$jmaiz/software. This research has made use of the NASA/IPAC Extragalactic Database (NED) which is operated by the Jet Propulsion Laboratory, California Institute of Technology, under contract with the National Aeronautics and Space Administration.

This work has been supported by the Spanish Ministry of
Education and Science, under grant BES-2003-0852, project AYA2002-01055. MG-M is supported by the German federal department for education and research (BMBF) under the project numbers: 50OS0502 \& 50OS0801. 
\end{acknowledgements}





\begin{thebibliography}{}

\bibitem{AAH0009}
Alonso-Herrero, A. et al.\, 2009, A\&A, submitted
\bibitem{AAH0000}
Alonso-Herrero, A., Rieke, G. H., Rieke, M. J., Colina, L., Pérez-González, P. G. \& Ryder, S. D. 2006, ApJ, 650, 835
\bibitem{b11000}
Arribas, S., Colina, L., Monreal-Ibero, A., Alfonso, J., Garc\'{\i}a-Mar\'{\i}n, M. \& Alonso-Herrero, A. 2008, A\&A, 479, 687
\bibitem{Ar2003}
Arribas, S., Colina, L., 2003, ASP Conference Proceedings, 297, 24-28
\bibitem{b2}
Arribas, S. et al.\,1998, Proc. SPIE, 3355, 821
\bibitem{b3}
Arribas, S., Mediavilla, E., Garc\'{\i}a-Lorenzo, B., \& del Burgo, C. 1997, ApJ, 490, 227
\bibitem{Bed2009}
Bedregal, A. G., Colina, L., Alonso-Herrero, A. \& Arribas, S. 2009, arXiv0904.3324
\bibitem{b5}
Bingham, R. G., Gellatly, D. W., Jenkins, C. R., \& Worswick, S. P. 1994, Proc. SPIE, 2198, 56
\bibitem{b550000}
Bushouse, H. A., Borne, K. D., Colina, L., Lucas, R. A., Rowan-Robinson, M., Baker, A. C., Clements, D. L., Lawrence, A. \& Oliver, S. 2002, ApJS, 138, 1
\bibitem{b5511100}
Colina, L., Arribas, S. \& Monreal-Ibero, A. 2005, ApJ, 621, 725C
\bibitem{b5500}
Colina, L., Borne, K., Bushouse, H., Lucas, R. A., Rowan-Robinson, M., Lawrence, A., Clements, D., Baker, A., \& Oliver, S. 2001, ApJ, 563, 546
\bibitem{b5500c}
Colina, L., Arribas, S., Borne, K. D. \& Monreal, A. 2000, ApJ, 533, L9
\bibitem{b550000c}
Colina, L., Arribas, S. \& Borne, K. D 1999, ApJ, 527L, 13
\bibitem{F01}
Farrah, D., Rowan-Robinson, M., Oliver, S., Serjeant, S., Borne, K., Lawrence, A., Lucas, R. A., Bushouse, H., \& Colina, L. 2001, MNRAS, 326, 1333
\bibitem{maca3}
Garc\'{\i}a-Mar\'{\i}n , M.,Colina, L., Arribas, S. \& Monreal-Ibero A. 2009, A\&A, accepted
\bibitem{maca2}
Garc\'{\i}a-Mar\'{\i}n , M.,Colina, L., Arribas, S., Alonso-Herrero, A. \& Mediavilla, E. 2006, \apj, 650, 850
\bibitem{b100}
Genzel, R., Tacconi, L. J., Rigopoulou, D., Lutz, D., \& Tecza, M. 2001, ApJ, 563, 527
\bibitem{b1112}
Howarth, I. D., \& Murray, J. 1988, DIPSO A Friendly Spectrum Analysis Program (Starlink User Note 50; Chilton: Rutherford Appleton Lab.)
\bibitem{Im111223}
Imanishi, M., Dudley, C. C., Maiolino, R., Maloney, P. R., Nakagawa, T. \& Risaliti, G. 2007, ApJS, 171, 72
\bibitem{Ken98}
Kennicutt, Robert C., Jr. 1998, ARA\&A, 36, 189
\bibitem{b30}
Leitherer, C., Schaerer, D., Goldader, J. D., Delgado, R. M., Robert, C., Kune, D. F., de Mello, D. F., Devost, D., \& Heckman, T. M. 1999, ApJS, 123, 3
\bibitem{b1222555}
LeFevre, O. et al. 2003, SPIE, 4841, 1670
\bibitem{b1222}
Lonsdale, C., Farrah, D. \& Smith, H. 2006, Astrophysics Update 2, edited by John W. Mason. ISBN 3-540-30312-X. Published by Springer Verlag, Heidelberg, Germany, 285
\bibitem{b122}
Low, J., Kleinmann, \& D. E. 1968, AJ, 73, 868
\bibitem{Monreal07}
Monreal-Ibero, A., Colina, L., Arribas, S. \& Garc\'{\i}a-Mar\'{\i}n, M. 2007, A\&A, 472, 421
\bibitem{Monreal06}
Monreal-Ibero, A., Arribas, S. \& Colina, L. 2006, ApJ, 637, 138
\bibitem{b1788}
Murphy, T. W., Jr., Soifer, B. T., Matthews, K., Armus, L. \& Kiger, J. R. 2001, AJ, 121, 97
\bibitem{b17}
Naab, T., Jesseit, R. \& Burkert, A. 2006, MNRAS, 372, 839
\bibitem{b17712}
Nardini, E., Risaliti, G., Salvati, M., Sani, E., Imanishi, M., Marconi, A. \& Maiolino, R. 2008, MNRAS, 385L, 130
\bibitem{b18}
Origlia L., \& Leitherer C. 2000, AJ, 119, 2018
\bibitem{b177}
Rieke, G. H. \& Low, F. J. 1972, ApJ, 176L, 95
\bibitem{Risa06}
Risaliti, G., Maiolino, R., Marconi, A., Sani, E., Berta, S., Braito, V., Ceca, R. D., Franceschini, A., \& Salvati, M. 2006, MNRAS, 365, 303 
\bibitem{Roth000}
Roth, M. et al.\, 2005, PASP, 117, 620
\bibitem{b20}
Sanders, D. B. \& Mirabel, I. F. 1996, ARA\&A, 34, 749
\bibitem{b4300}
Savage, B. D., \& Mathis, J. S. 1979, ARA\&A, 17, 73
\bibitem{b2002}
Scoville, N. Z., Evans, A. S., Thompson, R., Rieke, M., Hines, D. C., Low, F. J., Dinshaw, N., Surace, J. A. \& Armus, L. 2000, AJ, 119, 991
\bibitem{So084}
Soifer, B. T., Rowan-Robinson, M., Houck, J. R., de Jong, T., Neugebauer, G., Aumann, H. H., Beichman, C. A., Boggess, N., Clegg, P. E., Emerson, J. P. \& 6 coauthors 1984, AjP, 278L, 71
\bibitem{b21}
Surace, J. A., Sanders, D. B., Vacca, W. D., Veilleux, S. \& Mazzarella, J. M. 1998, ApJ, 492, 116
\bibitem[2002]{Tac02}
Tacconi, L. J., Genzel, R., Lutz, D., Rigopoulou, D., Baker, A. J., Iserlohe, C., \& Tecza, M. 2002, ApJ, 580, 73
\bibitem{b24}
Veilleux, S., Kim, D.C., Sanders, D. B., Mazzarella, J. M., \& Soifer, B. T. 1995, ApJS, 98, 171

\end{thebibliography}
\end{document}